\documentclass[pra,aps,epsf,nofootinbib,twocolumn,
notitlepage,showpacs,10pt]{revtex4-1}

\usepackage{color}
\usepackage{float}
\usepackage{graphicx}
\usepackage{subcaption}
\usepackage{amsmath}
\usepackage{amsthm}
\usepackage{mathtools}

\theoremstyle{definition}

\usepackage{bm}
\usepackage{hyperref}
\usepackage[normalem]{ulem}
\usepackage{float}
\restylefloat{table}

\usepackage{lipsum}

\theoremstyle{definition}

\usepackage{graphicx} %to insert graphs
\graphicspath{ {rrImages/} } %directory name
\usepackage{tikz} %to draw graphs
\usetikzlibrary{matrix,positioning}
\usepackage{amsfonts}
\usepackage{amssymb}
\theoremstyle{definition}
\newtheorem{definition}{Definition}[section]
\hypersetup{
	colorlinks = true,
	urlcolor=blue
}
\makeatletter
\def\thickhline{%
	\noalign{\ifnum0=`}\fi\hrule \@height \thickarrayrulewidth \futurelet
	\reserved@a\@xthickhline}
\def\@xthickhline{\ifx\reserved@a\thickhline
	\vskip\doublerulesep
	\vskip-\thickarrayrulewidth
	\fi
	\ifnum0=`{\fi}}
\makeatother

\newlength{\thickarrayrulewidth}
\begin{document}
	
	\markboth{James Ostrowski}
	{Impact of Graph Structures for QAOA on MaxCut}
	
	\title{Impact of Graph Structures for QAOA on MaxCut}
	
	\author{Rebekah Herrman}
	\email{rherrma2@tennessee.edu}
	\affiliation{
		Department of Industrial and Systems Engineering, The University of Tennessee\\Knoxville, Tennessee  37996-2315 USA}
	
	\author{Lorna Treffert}
	%\email{rherrma2@utk.edu}
	\affiliation{
		Department of Industrial and Systems Engineering, The University of Tennessee\\Knoxville, Tennessee  37996-2315 USA}
	
	\author{James Ostrowski}
	\email{jostrows@tennessee.edu}
	\affiliation{
		Department of Industrial and Systems Engineering, The University of Tennessee\\Knoxville, Tennessee  37996-2315 USA}
	\author{Phillip C. Lotshaw}
	%\email{lotshawpc@ornl.gov}
	\affiliation{
		Quantum Computing Institute\\ Oak Ridge National Laboratory\\ Oak Ridge, Tennessee 37830 USA}
	
	\author{Travis S. Humble}
	%\email{humlets@ornl.gov}
	\affiliation{
		Quantum Computing Institute\\ Oak Ridge National Laboratory\\ Oak Ridge, Tennessee 37830 USA}

	\author{George Siopsis}
	\email{siopsis@tennessee.edu}
	\affiliation{
		Department of Physics and Astronomy, The University of Tennessee\\Knoxville, Tennessee 37996-1200 USA}
	
	\begin{abstract}
		The quantum approximate optimization algorithm (QAOA) is a promising method of solving combinatorial optimization problems using quantum computing. QAOA on the MaxCut problem has been studied extensively on specific families of graphs, however, little is known about the algorithm on arbitrary graphs.  We evaluate the performance of QAOA at depths at most three on the MaxCut problem for all connected non-isomorphic graphs with at most eight vertices and analyze  how graph structure affects QAOA performance. Some of the strongest predictors of QAOA success are the existence of odd-cycles and the amount of symmetry in the graph. The data generated from these studies are shared in a publicly-accessible database to serve as a benchmark for QAOA calculations and experiments. Knowing the relationship between structure and performance can allow us to identify classes of combinatorial problems that are likely to exhibit a quantum advantage.

	\end{abstract}
	
	\maketitle

	%%%%%%%%%%%%%%%%%%%%%
	%%%%%%%%%%%%%%%%%%%%%
	
	\section{Introduction}\label{intro}
	
	Quantum computing has been shown to provide a theoretical advantage over classical computing in different areas such as machine learning and algorithms on shallow circuits \cite{bravyi2018quantum, riste2017demonstration}, and noisy intermediate-scale quantum (NISQ) devices provide an opportunity to test this advantage. The quantum approximate optimization algorithm (QAOA) is a promising application of NISQ devices developed by Farhi, Goldstone, and Gutmann to solve combinatorial optimization problems \cite{farhi2014quantum}. Unfortunately, large scale testing of QAOA has been difficult, as the size of problems that can be tested on NISQ hardware is limited and quantum simulations can be time consuming for small problems. 
	
	The current literature has mostly focused on examining graphs with a predetermined structure. The impact of problem structure on computational efficiency is well understood in optimization community~\cite{conforti2014integer}: it is common for a difficult class of problems to become easy when a certain structure is imposed. By  focusing on graphs with a predetermined structure in QAOA, there is a risk that conclusions made will not extend to the broader class of problems. In this paper, we seek to remedy this issue by performing an exhaustive analysis of how graph structure can impact QAOA on small MaxCut problems.

	While there are numerous classes of combinatorial optimization problems, the MaxCut problem has been the major focus of quantum computing research.  The MaxCut problem on an undirected, simple graph $G=(V(G),E(G))$ is to determine how to partition $V(G)$ into two sets such that the number of edges between them is maximized. This problem is classically hard, however, heuristics exist that find near optimal solutions \cite{festa2002randomized, goemans1994879}.
	
	%The QAOA algorithm consists of applying two operators,  $e^{-i C \gamma}$ and $e^{-i B \beta}$ alternatingly $p$ times each for level $p$ QAOA to an initial state dependent on choice of $B$. When we say $p=0$, we are referring to the initial state with no operators acting on it, so there is an equal probability of measuring any quantum state. Typically, $B$ is the sum of the Pauli- x operator acting on each qubit, although other mixers may be chosen, and $C$ depends on the optimization problem to be solved. For MaxCut in particular, $C= \sum_{uv \in E(G)} C_{uv}$ where $C_{uv} = \sigma_u^z \otimes \sigma_v^z$. %Thus, if QAOA is applied $p$ times to an initial state $\ket{s}$, the final state of the system is
	%\begin{equation*}
	%  \ket{\psi} =( \prod_{j=0}^{p-1} e^{-iB\beta_j}e^{-iC\gamma_j}) \ket{s}.
	%\end{equation*}
	
	QAOA for MaxCut has been studied on two and three regular graphs \cite{farhi2014quantum, brandao2018concentration, Medvidovic2020QAOA54qubit, zhou2018quantum, akshay2020reachability}, and random regular bipartite graphs \cite{farhi2020quantumwholegraph}, which are families that are not representative of the vast majority of graphs. Other work has focused on more general graphs, for instance Erd\H{o}s-R\'enyi random graphs \cite{farhi2020quantum}, but over small samples. Additionally, some of the more commonly studied measure of QAOA success are the expected value of $C$, $\langle C \rangle$, and the approximation ratio \cite{ozaeta2020expectation}, which is $\frac{\langle C \rangle}{C_{max}}$, where $C_{max}$ is the size of the cut in an optimal solution. Our goal is to look at general graphs and determine which structures correlate to better performances of QAOA on MaxCut for not only $\langle C\rangle$ and $\frac{\langle C \rangle}{C_{max}}$ but also the probability of measuring an optimal solution and the change in the approximation ratio from level $p$ QAOA to level $p+1$.
	
	To this end, we analyze graph structures and compare them to simulations of the algorithm on MaxCut at one, two, and three levels for all non-isomorphic, connected graphs on up to eight vertices. Additionally, we examine correlations between the structures and the probability of obtaining an optimal solution with QAOA. We also look at correlations between the expected value of the cost operator $C$, the ratio of $\langle C \rangle$ to the optimal solution, and the percent difference in these ratios for consecutive iterations of QAOA in order to determine which problems are solved well on quantum computers. A summary of the results is found in Table \ref{tab:summary}. % The data we collected suggests bipartite, Eulerian, and distance regular graphs give higher probabilities of measuring the optimal solution than graphs not satisfying those properties on the same number of vertices at the same number of iterations for $p \geq 2$. Additionally, graphs that have a lot of edges, a small diameter, large clique size, few cut vertices, and lots of small, odd cycles have a higher correlation with $\langle C \rangle$. Thus, we expect that QAOA for MaxCut on highly symmetric, sparse graphs will have a relatively high probability of measuring the optimal solution.  
	Finally, we created a database containing all connected graphs on at most eight vertices and the structures that were examined. The details for accessing the database is found in Appendix \ref{database}.
	
	This paper is organized as follows. %First, we review MaxCut and QAOA in Sec. \ref{background}. 
	In Sec. \ref{graphpatterns}, we discuss graph characteristics and their correlations to QAOA performance metrics. We then summarize the results and discuss future work in Sec. \ref{conclusion}.

	\section{Graph Structures}\label{graphpatterns}
	
	In previous work, we ran numerical simulations of QAOA solving the MaxCut problem on all non-isomorphic graphs with the number of vertices, $n$, ranging between three and eight \cite{QAOABFGS,lotshawdataset}. The angles that maximize $\langle C \rangle$ after one iteration of QAOA were solved using the open source software Couenne\cite{belotti2009couenne}.  For larger $p$, we used the angles, values of $\langle C \rangle$ and probabilities of measuring an optimal solution found in \cite{QAOABFGS}. The BFGS algorithm \cite{NumericalRecipesBFGS} was used as a heuristic to determine optimized angles using hundreds of random seeds for each graph. The optimized results are consistent between different implementations and are confirmed to be global optimal solutions in small cases with $n \leq 6$ and $p=2$ by comparing against a brute force method.  Thus, the correlations we draw can be thought of as applying to ``best-case" results for QAOA.  Using the numerical simulations from earlier work, we determine QAOA metrics of interest. The metrics used are:
	\begin{itemize}
		\itemsep0em 
		\item $\langle C \rangle$: The expected value of $C$
		\item $P(C_{max})$: The probability of measuring a state that represents a maximum cut
		\item Level $p$ approximation ratio: $\frac{\langle C \rangle_p}{C_{max}}$
		\item Percent change in approximation ratio ($\Delta$ ratio) at $p$: $\frac{\frac{\langle C \rangle_p}{C_{max}} - \frac{\langle C \rangle_{p-1}}{C_{max}}}{1-\frac{\langle C \rangle_{p-1}}{C_{max}}} = \frac{\langle C \rangle_p - \langle C \rangle_{p-1}}{C_{max}-\langle C \rangle_{p-1}}$
	\end{itemize}

	\noindent where $\langle C \rangle_{p}$ is the expected value of $C$ after $p$ iterations of QAOA. %The expected value of $C$ can be written as the sum of expected values for the operator acting on each edge in the graph, so more iterations of QAOA allows the algorithm to ``see" more of the graph structure.
	$\langle C \rangle$ can be written as the sum of expected values for the operator acting on each edge in the graph, where each edge expectation value at $p$ iterations depends on vertices up to $p$ edges away, so more iterations of QAOA allows the algorithm to “see" more of the graph structure. The level $p$ approximation ratio is the expected value of the cost function for MaxCut for QAOA performed at level $p$ as a percentage of the optimal solution, and the $\Delta$ ratio tells us the rate of change of the level per iteration.
	
	For three vertex graphs, QAOA was only run on two levels, as the correct solution was found with two iterations. Larger graphs were  run on three levels. Table \ref{tab:number} lists the number of connected, non-isomorphic graphs on $n$ vertices. As there are only two connected graphs on three vertices, the data for $n=3$ does not give much insight into the correlations between the structures and different QAOA properties, so we exclude the data for three vertex graphs from all tables. We study only connected graphs because if the graph is disconnected, each component has fewer vertices, and we can independently solve each subproblem. We also study non-isomorphic graphs, as isomorphic graphs have the same properties and result in the same solutions.
	
	\begin{table}\footnotesize
		\centering
		\begin{tabular}{|c|c|}
			\hline
			%$n$ & $N_n$\\
			$n$ & Number of Connected, Non-isomorphic Graphs on $n$ vertices, $N_n$\\
			
			\hline
			3 & 2  \\
			\hline
			4 & 6\\
			\hline
			5 & 21\\
			\hline
			6 & 112\\
			\hline
			7 & 853 \\
			\hline
			8 & 11117 \\
			\hline
		\end{tabular}
		\caption{The number of connected, non-isomorphic graphs on $n$ vertices for $3 \leq n \leq 8$. %$N_n$ is the number of connected, non-isomorphic graphs on $n$ vertices.
		}
		\label{tab:number}
	\end{table}
	
	\begin{table}\footnotesize
		
		\centering
		\begin{tabular}{|c|c|c|c|c|}
			\hline
			Graph Property & $\langle C \rangle$ & $P(C_{max})$ & $\frac{\langle C \rangle_p}{C_{max}}$ & $\frac{\langle C \rangle_p - \langle C \rangle_{p-1}}{C_{max}-\langle C \rangle_{p-1}}$\\
			\hline
			%3:0 &  &  &  &  &  &  &  &  &  & \\
			%\hline
			%3:1 &  1 & -1 & 1 & 1 & -1 & -1 & -1 & 1 & 1 & -1\\
			%\hline
			%3:2 &  .981 & -.981 & .981 & .981 & -.981 & -.981 & -.981 & .981 & .981 & -.981\\
			%\hline
			%3:2 & \rh{NEED} & \rh{NEED} & \rh{NEED} & \rh{NEED} & \rh{NEED} & \rh{NEED} & \rh{NEED} & N/A & \rh{NEED} & \rh{NEED}\\
			%\hline
			Edges & = & - & - & = \\
			\hline
			Diameter & = &  & - & = \\
			\hline
			Clique number & - &  - &  - &  \\
			\hline
			Bipartite & - &  & - &   \\
			\hline
			Eulerian & = &  & = &  \\
			\hline
			Distance regular & = & = &  = &  \\
			\hline
			Number of cut vertices & = & = & +  & =  \\
			\hline
			Number of minimal odd cycles & - & - &  - &  - \\
			\hline
			Group size & = & = & = &  \\
			\hline
			Number of orbits & = & - & -  &  - \\
			\hline
		\end{tabular}
		\caption{The trends in correlations between the group properties and each QAOA metric. The symbol ``+" refers to correlations that tend to increase, ``-" refers to a correlation that tend to decrease, and ``=" refers to correlations that tend to stay constant for fixed $n$ as $p$ increases. An empty space refers to a property with no strong correlation or discernible trend.}
		\label{tab:summary}
	\end{table}
	
	We collected properties of the tested graphs and found the Pearson's correlation coefficient between them and each  metric. The properties examined were:
	
	\begin{itemize}
		\itemsep0em 
		\item Number of edges
		\item Diameter
		\item Clique number
		\item Number of cut vertices
		\item Number of minimal odd cycles
		\item Group size
		\item Number of orbits
		\item Bipartite (boolean)
		\item Distance regularity (boolean)
		\item Eulerian (boolean)
		
	\end{itemize}
	
	\begin{figure*}
		\centering
		\begin{subfigure}{.45\textwidth}
			\centering
			\includegraphics[width=.8\linewidth]{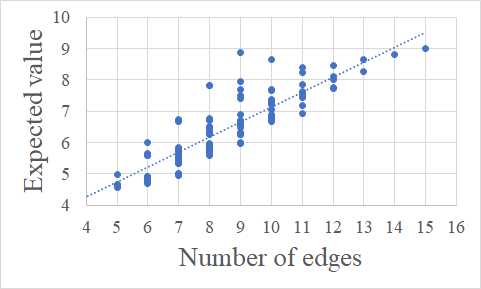}
			\caption{}
			\label{fig:edgesub1}
		\end{subfigure}%
		\begin{subfigure}{.45\textwidth}
			\centering
			\includegraphics[width=.8\linewidth]{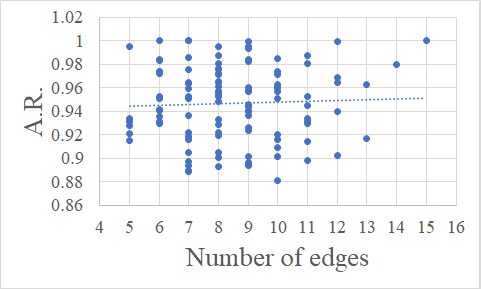}
			\caption{}
			\label{fig:edgesub2}
		\end{subfigure}
		
		\caption{$\langle C \rangle$ and the approximation ratio, denoted ``A.R.", plotted against the total number of edges in the graph for $p=3$. There is a strong positive correlation, with $r = .875$ in \ref{fig:edgesub1} and lack of correlation in \ref{fig:edgesub2}, with coefficient $r=.042$.}
		\label{fig:edgehistograms}
	\end{figure*}
	
	\begin{figure*}
		\centering
		\begin{subfigure}{.45\textwidth}
			\centering
			\includegraphics[width=.8\linewidth]{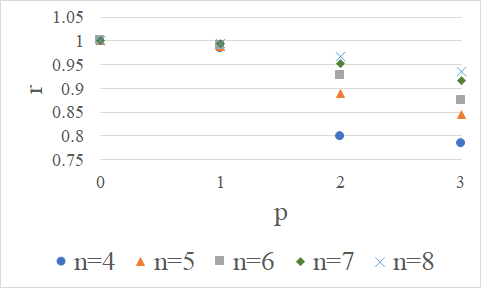}
			\caption{Correlation with $\langle C \rangle$}
			\label{fig:edgesub1}
		\end{subfigure}%
		\begin{subfigure}{.45\textwidth}
			\centering
			\includegraphics[width=.8\linewidth]{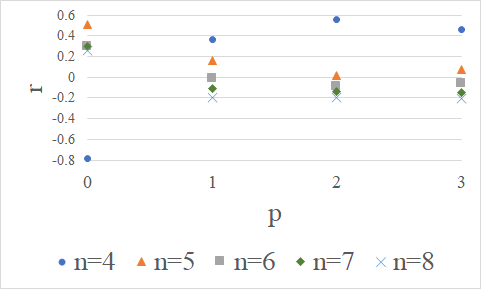}
			\caption{Correlation with $P(C_{max})$}
			\label{fig:edgesub2}
		\end{subfigure}
		\begin{subfigure}{.45\textwidth}
			\centering
			\includegraphics[width=.8\linewidth]{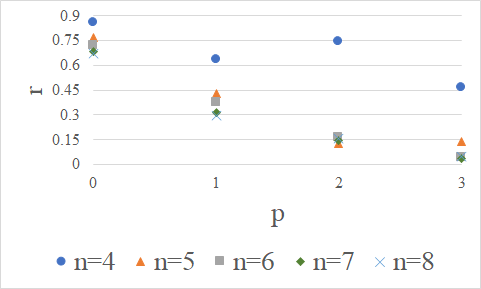}
			\caption{Correlation with $\frac{\langle C \rangle_p}{C_{max}}$}
			\label{fig:edgesub3}
		\end{subfigure}
		\begin{subfigure}{.45\textwidth}
			\centering
			\includegraphics[width=.8\linewidth]{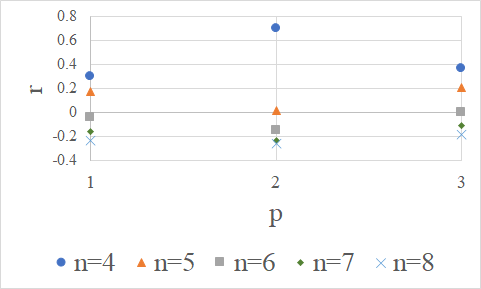}
			\caption{Correlation with $\frac{\langle C \rangle_p-\langle C \rangle_{p-1}}{C_{max}-\langle C \rangle_{p-1}}$}
			\label{fig:edgesub4}
		\end{subfigure}
		\caption{The correlations between the number of edges for each $n$ vertex graph and each of the QAOA metrics for $p$.}
		\label{fig:edgecorrelations}
	\end{figure*}
	
	Throughout the text and charts, $r$ denotes the correlation coefficient and $n$ denotes the number of vertices. Definitions of the graph theory terms are presented in Appendix \ref{definitions}. These properties are a subset of all the possible graph properties, but were chosen because they are commonly studied. We are providing the database to enable others to identify other structures of interest. 
	The tables listing correlation coefficients are found in Appendix \ref{correlations}. The graphs and numbering system are from the connected graph files by Brendan McKay \cite{mckay}. In the following subsections, we discuss graph properties and their correlations with different QAOA metrics. Throughout each section, we will focus on the behavior of $r$ as $n$ and $p$ increase, as these are small graphs that would not be representative of interesting problems that can provide a quantum advantage. Additionally, there are few graphs on four and five vertices compared to $n \in \{6,7,8\}$, so data points for $n \in \{4,5\}$ may not be representative of the trends in correlations. Thus, we will focus our analysis on $n \geq 6$. 
	
	\subsection{Edges, clique number, and minimal odd cycles}\label{edges section}
	The correlation between number of edges and the metrics computed varies greatly. There is a very strong correlation between the number of edges and $\langle C \rangle$, which can be seen in Fig. \ref{fig:edgehistograms} and Fig. \ref{fig:edgesub1}. This makes intuitive sense for the following reason. Arbitrarily adding edges to a given graph will not decrease the objective value of any given solution, only potentially increase it, so we would expect that denser graphs will have higher $\langle C \rangle$ values. However, the results in Fig. \ref{fig:edgesub2} shows that the number of edges is negatively correlated with the probability of sampling an optimal solution for large $n$ and $p$.  We believe this is in part because the more edges that are added in a graph, the more near optimal solutions there are that QAOA may favor over the optimal solution. These near-optimal solutions compete with exact optimal solutions in sampling, as QAOA optimizes $\langle C \rangle$, it can favor near optimal states over truly optimal ones. 
	
	Note that as $n$ increases, the correlation coefficient between edges and $\langle C \rangle$ increases towards $1$ for fixed $p$, while the coefficients with  $P(C_{max})$, the approximation ratio, and the $\Delta$ ratio slightly decrease, as seen in Fig. \ref{fig:edgecorrelations}. It is unclear why the size of $n$ impacts the correlation coefficient. At large $p$, we would expect that $\langle C \rangle$ would take the value of $C_{max}$, so eventually, all of the approximation ratios will reach one, making them independent of the number of edges. The negative correlation in the change in approximation ratios can be explained as follows. As previously mentioned, denser graphs have high $\langle C \rangle$, and thus tend to have good approximation ratios. Hence, there is less room for improvement, whereas sparser graphs have more.  We expect that for a fixed $n \geq 6$, $\Delta$ will approach zero as the number of iterations increases, based on Fig. \ref{fig:edgesub4}. The correlation is determined by the values of $\Delta$ and the fluctuations in $\Delta$ between graphs with various numbers of edges.  When $\Delta$ becomes small, the fluctuations between graphs overwhelm the typical variations in $\Delta$ with the number of edges, so the correlation tends to zero. % The numerator of $\Delta$ will eventually become zero, since the expected value cannot go above $C_{max}$. If it reaches $C_{max}$ at level $p$, it is still the same at all subsequent iterations.  %If the numerator tends to zero for each graph, the correlation will disappear. 
	\begin{figure}
		\centering
		\includegraphics[scale=.6]{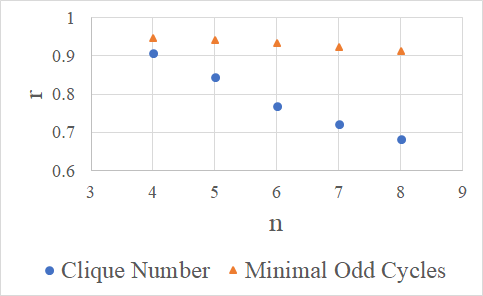}
		\caption{The correlation coefficients of number of edges with clique number and number of minimal odd cycles.}
		\label{correlationtwo}
	\end{figure}
	
	Since the edge density, clique number, and number of minimal odd cycles are correlated for small $n$, as seen in Fig. \ref{correlationtwo}, it is no surprise that they have strong correlations with the same QAOA metrics. Refer to Tables \ref{tab:prob}-\ref{tab:changeingap} to view their correlation coefficients. The properties become less correlated as $n$ increases so it is unclear if the properties will have similar $r$ values for higher vertex graphs. The properties may need separate analyses for larger graphs. It could be that even though the properties become less correlated with each other, they remain highly correlated with some of the QAOA metrics.
	\begin{figure*}
		\centering
		\begin{subfigure}{.45\textwidth}
			\centering
			\includegraphics[width=.8\linewidth]{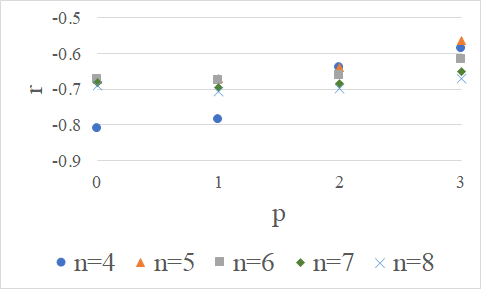}
			\caption{Correlation with $\langle C \rangle$}
			\label{fig:diamsub12}
		\end{subfigure}%
		\begin{subfigure}{.45\textwidth}
			\centering
			\includegraphics[width=.8\linewidth]{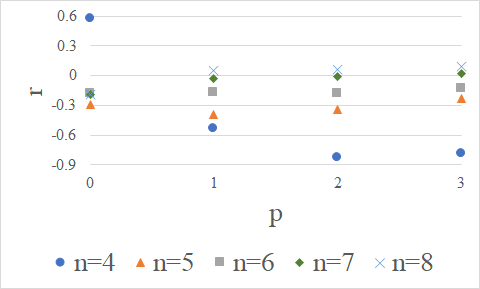}
			\caption{Correlation with $P(C_{max})$}
			\label{fig:diamsub22}
		\end{subfigure}
		\begin{subfigure}{.45\textwidth}
			\centering
			\includegraphics[width=.8\linewidth]{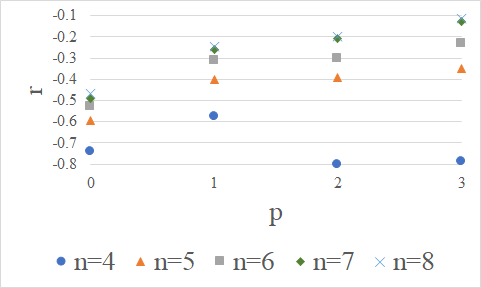}
			\caption{Correlation with $\frac{\langle C \rangle_p}{C_{max}}$}
			\label{fig:diamsub32}
		\end{subfigure}
		\begin{subfigure}{.45\textwidth}
			\centering
			\includegraphics[width=.8\linewidth]{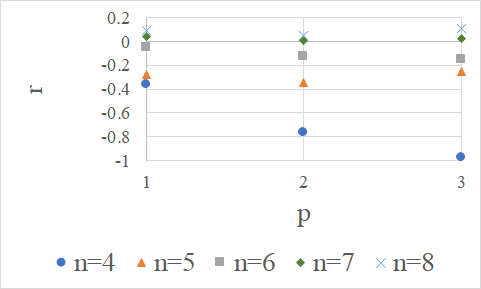}
			\caption{Correlation with $\frac{\langle C \rangle_p-\langle C \rangle_{p-1}}{C_{max}-\langle C \rangle_{p-1}}$}
			\label{fig:diamsub42}
		\end{subfigure}
		\caption{The correlations between the diameter for each $n$ vertex graph and each of the QAOA metrics for $p$.}
		\label{fig:diametercorrelations2}
	\end{figure*}
	
	%\begin{figure}
	%\centering
	%\includegraphics[scale=.6]{averageprobdiameter.png}
	%\caption{The average probability of measuring an optimal solution as a function of the diameter for $n=6$.}
	%\label{probdiam}
	%\end{figure}
	\begin{figure}
		\centering
		\includegraphics[scale=.6]{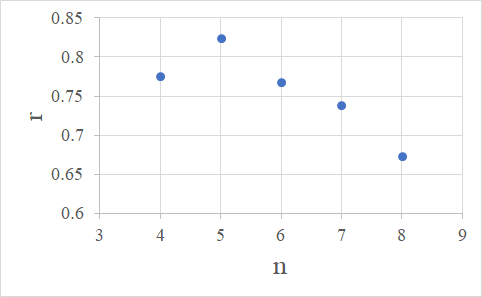}
		\caption{The correlation coefficients between diameter and number of cut vertices for graphs on $n$ vertices.}
		\label{diamcutvertices}
	\end{figure}

	In summary, the highest correlation between edges and the QAOA metrics is with $\langle C \rangle$. Thus, if a high expected value is the desired metric for success, graphs that are dense are optimal for QAOA. Edge density does not appear to have a large impact on $P(C_{max})$ and the approximation ratio, however simulations of larger graphs are needed to determine if the correlations between edges and these metrics tends towards a limit between zero and negative one, or instead continues towards it. If the limit approaches negative one, then graphs with low edge density may yield higher probabilities of measuring an optimal solution or obtaining a high approximation ratio.
	
	\subsection{Diameter and number of cut vertices}
	
	For the most part, the diameter of a graph, $d$, correlates negatively with the QAOA metrics, but the correlation becomes less negative as $n$ increases, which is the reverse of the correlations with the number of edges. The correlation is particularly strong with $\langle C \rangle$  for low values of $p$, as seen in Fig. \ref{fig:diamsub12}. This is not unexpected as diameter is negatively correlated with edge density. As seen in Fig. \ref{fig:diamsub22}, the correlation between diameter and probability of measuring an optimal state is stronger for smaller $n$, and tends to approach zero for fixed $n$ as $p$ increases. %We believe this is in part because when $p \geq \frac{d}{2}$, QAOA can see all the edges of the graph so the distance between any pair of edges is not important. 
	%It is worth noting that when $p$ increases from $\lceil \frac{d}{2} \rceil -1 $ to $ \lceil \frac{d}{2} \rceil$, the average probability of measuring an optimal solution does not dramatically increase compared to the increase from $\lceil \frac{d}{2} \rceil -2 $ to $\lceil \frac{d}{2} \rceil -1 $, as seen in Fig. \ref{probdiam}. 
	Similarly to the probability, the correlation between $d$ and the approximation ratio is stronger for smaller $n$, and tends to approach zero for fixed $n$ as $p$ increases, as does the correlation between $d$ and the $\Delta$ ratio. This is seen in Figs. \ref{fig:diamsub32} and \ref{fig:diamsub42}. 
	
	The number of cut vertices, like the diameter, correlates negatively with QAOA metrics, which makes sense as the two properties are positively correlated, as seen in Fig. \ref{diamcutvertices}, however the correlation weakens as $n$ increases. Thus, the two quantities may not correlate similarly with all of the QAOA metrics for larger $n$. % An explanation of the negative correlation between the number of cut vertices and QAOA performance is as follows. One can decompose a graph $G$ that contains a cut vertex, $v$, into smaller graphs that are the connected components of $G \setminus v$. We can then add $v$ back into each connected component and solve each connected component separately. The solutions for each subproblem can be pieced back together into a solution for the entire graph. If the $v$ is in different sets for each subproblem, we can swap the sets that each vertex belongs to in the subproblem at hand until they align. We expect that solving each subproblem individually gives better performance than solving the entire problem, since there are fewer interactions to consider. There is the possibility that each subproblem requires different angles to optimize the metrics, however previous work shows that most graphs are optimized with similar angles \cite{QAOABFGS}. 
	See Tables \ref{tab:prob}-\ref{tab:changeingap} to see how the correlation coefficients compare between diameter and number of cut vertices.
	
	As $n$ increases, the correlation coefficient between diameter and $\langle C \rangle$ decreases as $p$ increases, while the other metrics tend to increase, as seen in Fig. \ref{fig:diametercorrelations2}. In fact, the correlation with the other metrics approaches zero for high $n$ and $p$. If these trends continue for $n \geq 9$, it would imply that diameter is not indicative of the success QAOA has when solving a MaxCut problem.
	\begin{figure*}
		\centering
		\begin{subfigure}{.45\textwidth}
			\centering
			\includegraphics[width=.8\linewidth]{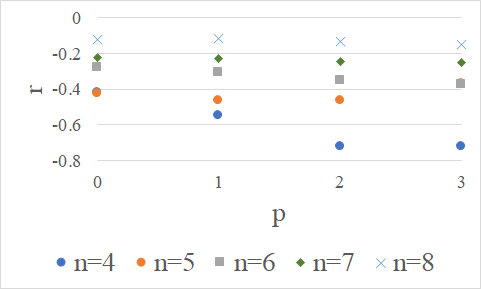}
			\caption{Correlation with $ \langle C \rangle$}
			\label{fig:orbitsub12}
		\end{subfigure}%
		\begin{subfigure}{.45\textwidth}
			\centering
			\includegraphics[width=.8\linewidth]{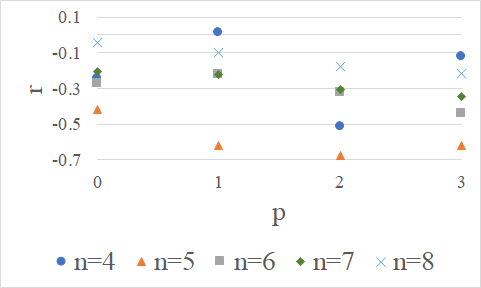}
			\caption{Correlation with $P(C_{max})$}
			\label{fig:orbitsub22}
		\end{subfigure}
		\begin{subfigure}{.45\textwidth}
			\centering
			\includegraphics[width=.8\linewidth]{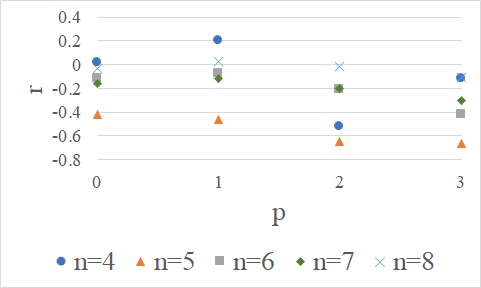}
			\caption{Correlation with $\frac{\langle C \rangle_p}{C_{max}}$}
			\label{fig:orbitsub32}
		\end{subfigure}
		\begin{subfigure}{.45\textwidth}
			\centering
			\includegraphics[width=.8\linewidth]{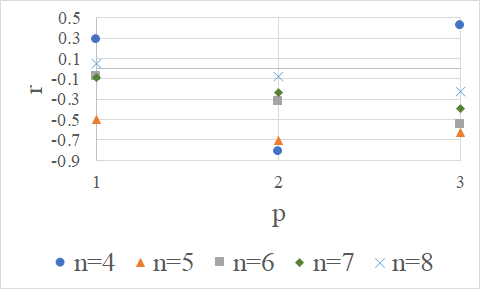}
			\caption{Correlation with $\frac{\langle C \rangle_p-\langle C \rangle_{p-1}}{C_{max}-\langle C \rangle_{p-1}}$}
			\label{fig:orbitsub42}
		\end{subfigure}
		\caption{The correlations between the number of orbits for each $n$ vertex graph and each of the QAOA metrics for $p$.}
		\label{fig:orbitcorrelations}
	\end{figure*}
	%These properties do not have a strong correlation with the probability of measuring an optimal solution, and the correlation between them and the approximation ratio weakens as $n$ and $p$ grow. Thus, diameter and the number of cut vertices do not seem to be properties that can predict the success of QAOA for MaxCut. 
	\subsection{Group Size}
	The group size of a graph tends to have high positive correlations with all QAOA properties for small $n$. However as $n$ increases, the correlations tend to zero. A zero correlation, however, might not indicate that the two are unrelated in this case. The size of the symmetry groups range from $1$ to $2^n$ for all graphs on $n$ vertices. Correlations are intended to measure linear relationships, so using data that is scaled exponentially on $n$ could ``wash out'' the correlations. Instead of group size, we look at the number or orbits to better investigate the relationship between symmetry and QAOA performance. 
	
	\subsection{Number of orbits}
	An automorphism of a graph is a relabeling of vertices that preserves edges, and therefore is a type of symmetry. An orbit of a vertex $v$ is the set of vertices with which $v$ can be swapped in an automorphism. Thus, if there are two vertices in the same orbit but in different sets of an optimal MaxCut solution, there is a possibility that they can be swapped to give another optimal solution. If there are fewer large orbits, there are more possible symmetries than with many small orbits.
	
	All vertices in the same orbit have the same degree, else the edges cannot be preserved. If there are more orbits, there are fewer symmetries, as there are fewer potential vertex mappings that preserve edges. It is not obvious why fewer orbits tends to produce better expected values and better approximation ratios.
	In Figs. \ref{fig:orbitsub12}, \ref{fig:orbitsub22}, \ref{fig:orbitsub32}, and \ref{fig:orbitsub42}, we see that, upon discarding $n=4$, the correlation coefficient between number of orbits and the QAOA metrics tends to increase as $n$ increases. Additionally, the correlation coefficient with the approximation ratio becomes more negative as $p$ increases. %This might mean that QAOA could be very effective on highly symmetric graphs.
	Hence, graphs with small group sizes and larger orbits should achieve higher approximation ratios after several iterations of QAOA. Thus, group size and symmetry should play an important role in the quality of solution, which has been noted by Shaydulin, Hadfield, Hogg, and Safro \cite{shaydulin2020symmetry}.
	
	\subsection{Bipartite}
\begin{figure}
		\centering
		\includegraphics[scale=.6]{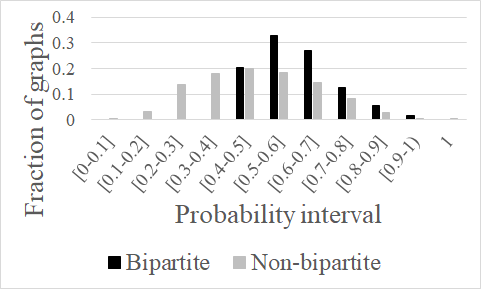}
		\caption{The fraction of eight vertex bipartite and non-bipartite graphs with a given probability of finding an optimal solution for $p=3$.
			%A histogram depicting the number of graphs with eight vertices whose probability of measuring an optimal solution at $p=3$ lies within each interval as a percentage of the number of bipartite and non-bipartite graphs. Note that non-bipartite graphs have probabilities between $0$ and $1$ while bipartite range from $0.4$ to $1$. The first and last columns appear empty, however they contain less than ten non-bipartite graphs each.
		}
		\label{bipartiteplot}
	\end{figure}

	Since there are so few bipartite graphs compared to non-bipartite for fixed $n$, looking at the average of each QAOA metric offers more insight into how a graph being bipartite affects the quality of QAOA solution. The average for all QAOA metrics are lower for bipartite graphs than non-bipartite when $p=1$ for fixed $n$, however, the probability and change in approximation ratio are higher for bipartite graphs for larger $p$ as seen in Table \ref{tab:bipartite} and Figure \ref{bipartiteplot}. Depending on the preferred metric of success, bipartite graphs either perform worse than non-bipartite, as in the case for the approximation ratio and $\langle C \rangle$, or better, as in the case of probability of measuring an optimal solution or $\Delta$ ratio. The average $\Delta$ ratio being higher for bipartite graphs suggests that a low approximation ratio at $p=1$ does not imply that the approximation ratio at larger iterations cannot surpass those of non-bipartite graphs. %Additionally, the data indicates that bipartite graphs tend to do better once QAOA has seen that there are no odd cycles. The data for bipartite graphs provides evidence that strong performance at $p=1$ does not imply strong performance for later iterations, as the change in approximation ratios is higher for bipartite graphs than non-bipartite in general, even though the approximation ratio tends to be higher for non-bipartite graphs. 
	The fact that the approximation ratio tends to be lower for bipartite graphs agrees with previous literature \cite{farhi2020quantumwholegraph}. In fact, Wurtz and Love conjecture that for $p$ iterations of QAOA on any graph, the approximation ratio of an $n$ vertex graph is lower when there are no odd cycles of length at most $2p+1$ than graphs that have length at most $2p+1$ \cite{wurtz2020bounds}. %For graphs with at most seven vertices, there are few odd cycles of lengths five and seven. 

	\subsection{Eulerian}
	\begin{figure}
		\centering
		\includegraphics[scale=.6]{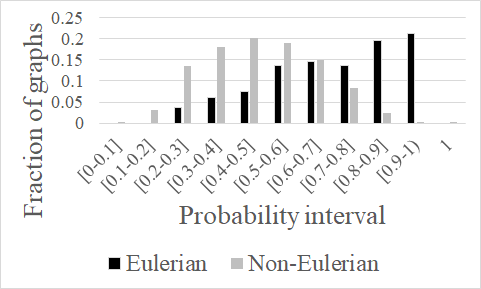}
		\caption{The fraction of eight vertex Eulerian and non-Eulerian graphs with a given probability of finding an optimal solution for $p=3$.
			%A histogram depicting the number of graphs with eight vertices whose probability of measuring an optimal solution at $p=3$ lies within each interval as a percentage of the number of Eulerian and non-Eulerian graphs. Note that non-Eulerian graphs have probabilities between $0$ and $1$ while Eulerian range from $0.2$ to $1$. The first and last bins appear empty, however they contain less than ten non-Eulerian graphs each.
		}
		\label{eulerianplot}
	\end{figure}
	
	There are so few Eulerian graphs in the data set that the correlation coefficients are not particularly helpful, however plotting the performance of Eulerian graphs makes the relationship between them and the QAOA metrics more evident.  As seen in Table \ref{tab:eulerian}, the average maximum probability, level $p$ approximation ratio, and $\Delta$ ratios are significantly higher for Eulerian graphs than for non-Eulerian for most $n$ and $p$, while the averages are comparable for the expected value of $C$. Figure \ref{eulerianplot} shows a histogram of the number of Eulerian and non-Eulerian graphs on eight vertices and the probability of obtaining an optimal solution. As seen in the histogram, for any given graph, the probability of measuring an optimal MaxCut solution is higher for Eulerian graphs than non-Eulerian.

	\subsection{Distance regular}
	Similarly to Eulerian graphs, distance regularity does not appear to have a strong correlation with any of the QAOA metrics because there are so few distance regular graphs compared to non-distance regular ones, so we do not make a chart of the correlations. In particular, there are only ten distance regular graphs on fewer than seven vertices. We also do not make a histogram with the probabilities as we did for Eulerian and bipartite graphs because the average probability of measuring an optimal solution for only a handful of distance regular graphs cannot be compared as well to the average probability of thousands of non-distance regular graphs. However, graphs with $n \in \{4,5\}$ vertices that are distance regular have a probability of one of obtaining the optimal solution when $p=2$. These graphs in particular are $C_4$, $K_4$, $C_5$ and $K_5$, where $C_n$ denotes the cycle on $n$ vertices and $K_m$ denotes the complete graph on $m$ vertices. These are the only distance regular graphs on four or five vertices, up to isomorphism. Interestingly enough, when $n=7$, both distance regular graphs, $C_7$ and $K_7$ obtain a probability of one of being in the optimal state when $p=3$, however, this is not the case for $n=6$. When $n=6$, three of the four distance regular graphs have a probability greater than $0.99$ of giving the correct solution, however the complete bipartite graph with partitions both containing three vertices ($K_{3,3}$) achieves a probability of $0.97$. This is far higher than the average for arbitrary graphs with $n=6$ and $p=3$.  These results lead us to believe that distance regular graphs on $n$ vertices achieve a high probability of obtaining the optimal solution when $p \geq \lfloor \frac{n}{2} \rfloor$, however more testing and a rigorous mathematical proof would be needed to confirm this.
	
	%\subsection{Group size and number of orbits}
	%Larger group sizes are usually, but not always, associated with a smaller number of orbits, so it is not surprising that the correlations between group size are opposite that of the orbit size. One caveat with group size, however, is that it has a range of 1 to $2^n$. This exponential scaling may wash out some of the effects. With that in mind, we focus primarily on number of orbits. 

	%Fewer orbits tends to imply that there are multiple optimal solutions, which is why the probability of finding {\em an} optimal solution is negatively correlated with the number of orbits. It is not obvious to us why fewer orbits tends to produce better expected values and better gaps, but note that the changes in the gap get stronger as the size of $p$ increases. This might mean that QAOA could be very effective on highly symmetric graphs. 

	%The group size correlates strongly with all properties for low $n$ and $p$, while the number of orbits has a strong negative correlation in similar instances. This is surprising because the group size and number of orbits are related to the symmetries of a graph, so it was expected that they would correlate highly with the expected value of $C$. One reason this does not occur is that the number of graphs with nontrivial group sizes or orbits may be small compared to the total number of graphs. 

	\section{Conclusion}\label{conclusion}
	The quantum approximate optimization algorithm has been studied in detail for the MaxCut problem on graphs with rigid structures including degree regularity and whether or not a graph is bipartite \cite{farhi2014quantum, brandao2018concentration, Medvidovic2020QAOA54qubit, zhou2018quantum}. While the studies give insight into how the algorithm works, examining graphs with specific structures excludes the vast majority of graphs. Thus, we looked at properties of all connected graphs on at most eight vertices and found the correlation between different graph properties and the probability of finding an optimal solution, the expected value of $C$, the level $p$ approximation ratio, and the change in ratio from $p-1$ to $p$ for $p$ at most three. 
	
	There are two metrics that strongly correspond with the graph properties studied, namely $\langle C \rangle$ and $P(C_{max})$. Graphs that have a lot of edges have a high positive correlation with with $\langle C \rangle$.  Diameter, clique size, number of cut vertices, and number of small odd cycles also correlate with $\langle C \rangle$, either positively or negatively. This is expected because these properties correlate positively or negatively with the number of edges for small $n$. Trends in the data for fixed $n$ and increasing $p$ are summarized in Table \ref{tab:summary}. The probability of measuring the optimal solution, expected value of $C$, and optimal angles were taken from  \cite{QAOABFGS,lotshawdataset}.
	
	Bipartite, Eulerian, and distance regular graphs tend to have higher probabilities of measuring the optimal solution than graphs that do not have any of these properties. For bipartite, this occurs for $p \geq 2$. Previous work shows that the approximation ratio for bipartite graphs tends to be worse than non-bipartite \cite{farhi2020quantumwholegraph}, so depending on the metric of success, bipartite graphs may or may not be suitable graphs for testing. Similarly, for larger $p$, Eulerian and distance regular graphs tend to have higher probabilities of measuring the optimal solution than non-Eulerian and non-distance regular graphs. Thus, we expect that QAOA for MaxCut on highly symmetric graphs, which are those with larger orbits, bipartite, Eulerian, and distance regular graphs will have a relatively high probability of measuring the optimal solution.  Table \ref{tab:summarytwo} gives a summary of the average correlation coefficient over all $p \geq 1$ for each graph property and each QAOA metric on graphs with eight vertices.
	
	\begin{table}\footnotesize
		\centering
		\begin{tabular}{|c|c|c|c|c|}
			\hline
			Graph Property & $\langle C \rangle$ & $P(C_{max})$ & $\frac{\langle C \rangle_p}{C_{max}}$ & $\frac{\langle C \rangle_p - \langle C \rangle_{p-1}}{C_{max}-\langle C \rangle_{p-1}}$\\
			\hline
			%3:0 &  &  &  &  &  &  &  &  &  & \\
			%\hline
			%3:1 &  1 & -1 & 1 & 1 & -1 & -1 & -1 & 1 & 1 & -1\\
			%\hline
			%3:2 &  .981 & -.981 & .981 & .981 & -.981 & -.981 & -.981 & .981 & .981 & -.981\\
			%\hline
			%3:2 & \rh{NEED} & \rh{NEED} & \rh{NEED} & \rh{NEED} & \rh{NEED} & \rh{NEED} & \rh{NEED} & N/A & \rh{NEED} & \rh{NEED}\\
			%\hline
			Edges & + & - & + & - \\
			\hline
			Diameter & - &  & - &  \\
			\hline
			Clique number & + & - & + & - \\
			\hline
			Bipartite & + &  & + &   \\
			\hline
			Eulerian &  & - &  & - \\
			\hline
			Distance regular &  &  &   &  \\
			\hline
			Number of cut vertices & - &  &   &   \\
			\hline
			Number of minimal odd cycles & + & - &  + & -  \\
			\hline
			Group size &  &  &  &  \\
			\hline
			Number of orbits & - & - &   &  \\
			\hline
		\end{tabular}
		\caption{A summary of graph properties and average correlation coefficient over all $p \geq 1$ to QAOA metrics for $n = 8$. A ``$+$" means a strong positive average coefficient, ``$-$" represents a strong average negative correlation coefficient, and an empty space means the average correlation coefficient is in the interval $(-.1,.1)$, which is not significant.}
		\label{tab:summarytwo}
	\end{table}
	
	Additionally, we created a dataset containing the graphs and their properties. The dataset information is located in Appendix \ref{database}. For future work, a similar data set for other problems, such as maximum independent set or problems with low circuit depth \cite{Herrman2020depth} would be useful to create benchmarks for experiments on NISQ devices. Other avenues of future work include using machine learning to determine if MaxCut on a graph will have a high probability of measuring the optimal solution after $p$ rounds of QAOA. Mathematically proving if QAOA for MaxCut on distance regular graphs with $n$ vertices gives better solutions than non distance regular graphs after $\lfloor \frac{n}{2} \rfloor$ iterations would be of interest, as well.
	
	\section*{Acknowledgements}
	The authors would like to thank Ryan Bennink for his helpful comments on early drafts of this manuscript.
	
	This work was supported by DARPA ONISQ program under award W911NF-20-2-0051. J. Ostrowski acknowledges the Air Force Office of Scientific Research award, AF-FA9550-19-1-0147. G. Siopsis acknowledges the Army Research Office award W911NF-19-1-0397. J. Ostrowski and G. Siopsis acknowledge the National Science Foundation award OMA-1937008.
	
	This manuscript has been authored by UT-Battelle, LLC under Contract No. DE-AC05-00OR22725 with the U.S. Department of Energy. The United States Government retains and the publisher, by accepting the article for publication, acknowledges that the United States Government retains a non-exclusive, paid-up, irrevocable, world-wide license to publish or reproduce the published form of this manuscript, or allow others to do so, for United States Government purposes. The Department of Energy will provide public access to these results of federally sponsored research in accordance with the DOE Public Access Plan. (http://energy.gov/downloads/doe-public-access-plan).

	\appendix
	
	\section{Graph Theory Definitions}\label{definitions}
	
	In this section, we define some of the less common graph theory terms that appear in the paper.
	
	\theoremstyle{definition}
	\begin{definition}[Diameter]
		The diameter of a graph $G=(V,E)$ is $\max_{u,v} d(u,v)$, where $d(u,v)$ denotes the distance between $u, v \in V$. 
	\end{definition}
	
	\theoremstyle{definition}
	\begin{definition}[Distance]
		The distance between two vertices, $u$ and $v$, of a graph $G=(V,E)$ is the number of edges in the shortest path between $u$ and $v$. 
	\end{definition}
	
	\theoremstyle{definition}
	\begin{definition}[Clique number]
		The clique number of a graph is the size of the largest complete subgraph.
	\end{definition}
	
	\theoremstyle{definition}
	\begin{definition}[Bipartite]
		A graph is bipartite if the vertices can be partitioned into two sets such that all edges are incident to a vertex in both sets.
	\end{definition}
	
	\theoremstyle{definition}
	\begin{definition}[Eulerian cycle]
		An Eulerian cycle is a cycle that uses all edges of a graph exactly once.
	\end{definition}
	
	\theoremstyle{definition}
	\begin{definition}[Eulerian]
		A graph is Eulerian if it is connected and contains an Eulerian cycle.
	\end{definition}
	
	\theoremstyle{definition}
	\begin{definition}[Distance Regular]
		A graph $G=(V,E)$ is distance regular if for any pair of vertices $x,y \in V$, the number of vertices that are distance $i$ from $x$ equals the number of vertices that are distance $i$ from $y$, for $i \in \{1, 2, ..., d\}$, where $d$ is the diameter of $G$. 
	\end{definition}
	
	\theoremstyle{definition}
	\begin{definition}[Cut vertex]
		A cut vertex of a connected graph is a vertex whose removal disconnects the graph.
	\end{definition}
	
	\theoremstyle{definition}
	\begin{definition}[Graph Automorphism]
		A graph automorphism is a relabeling of vertices that preserves the set of edges. Mathematically, it is a map $\alpha: V \longrightarrow V$ such that $ij \in E$ if and only if $\alpha(i)\alpha(j) \in E$. 
	\end{definition}
	
	\theoremstyle{definition}
	\begin{definition}[Automorphism Group]
		The set of all graph automorphisms of $G$ forms an automorphism group of $G$. 
	\end{definition}
	
	\theoremstyle{definition}
	\begin{definition}[Group Size]
		The group size of $G$ is the size of the automorphism group of $G$. 
	\end{definition}
	
	\theoremstyle{definition}
	\begin{definition}[Orbit]
		An orbit of a vertex $v$ is the set of all vertices $\alpha(v)$ where $\alpha$ is an automorphism of $G$. 
	\end{definition}

	\section{Database Details}\label{database}
	The data generated from these studies are shared in a publicly-accessible GitHub Repository to serve as a benchmark QAOA calculations and experiments. This data set consists of csv files for each set of graphs on $n$ vertices for $3 \leq n \leq 8$. The generated graph entries, which populate each set, are indexed by a graph number and include the following properties: bipartite (boolean), number of edges, diameter, clique number, distance regular (boolean), eulerian (boolean), list of cut vertices, number of cut vertices, cycle basis, degree sequence, automorphism group generator, automorphism group size, orbits, number of orbits, and the number of small cycles on 3 to $n$ vertices. In addition to the dataset, we have provided two options for storing and sorting this data by desired graph properties. The first option is to insert the data into a MySQL database structure. The user can then query the database to select the desired data and insert new columns for additional properties, calculations, and experimental results. We have provided scripts for creating the database tables, populating the tables using the csv files, querying the database, and inserting new columns. For additional information on how to download and use MySQL, see the \href{https://dev.mysql.com/doc/}{MySQL Documentation}.To work with the dataset directly, we have also included a python script which utilizes the data analysis library $pandas$. Pandas is a library which allows the user to store and manipulate data in 2-dimensional, labeled data structures called DataFrames. We have provided a python script which allows the user to import the csv data into pandas DataFrames and create new DataFrames with the desired graph properties. For more information on how to use and install pandas, see the \href{https://pandas.pydata.org/}{Pandas Documentation}. Both MySQL and Pandas are free, open-source software packages. This data set, the MySQL Scripts, and python script can be accessed through the 
	\href{https://github.com/jao204/QAOA_Small_Graph.git}{QAOA Small Graph GitHub Repository}.  
	
	\onecolumngrid
	
	\section{Data Tables}\label{correlations}
	In this section, we include the tables containing all correlations between different graph properties and metrics. For the boolean properties, we assign ``1" to TRUE and ``0" to FALSE.
	
	\begin{table}[H]\footnotesize
		\centering
		\begin{tabular}{|c|c|c|c|c|c|c|c|c|c|c|c|}
			\hline
			$n:p$ & Edges & Diam. & Clique num. & Bipartite & Eulerian  & Dist. reg. & Num. cut vertices & Num. min. odd cycles & Grp. size  & Num. orbits\\
			%\hline
			%3:0 &  &  &  &  &  &  &  &  &  & \\
			%    \hline
			%3:1 & 1 & -1 & 1 & 1 & -1 & -1 & -1 & 1 & 1 & -1\\
			%\hline
			% 3:2 & 1 & -1 & 1 & 1 & -1 & -1 & -1 & 1 & 1 & -1\\
			\hline
			%3:2 & \rh{NEED} & \rh{NEED} & \rh{NEED} & \rh{NEED} & \rh{NEED} & \rh{NEED} & \rh{NEED} & N/A & \rh{NEED} & \rh{NEED}\\
			%\hline
			4:0 & -.781 & .577 & -.894 & -1 & -.447 & 0 & .447 & -.866 & -.307 & -.243\\
			\hline
			4:1 & .363 & -.535 & .558 & .398 & .030 & .437 & -.085 & .346 & .507 & .015\\
			\hline
			4:2 & .561 & -.830 &  .421 & .247 & -.512 & .809 & -.769 & .354 & .663 & -.513 \\
			\hline
			4:3 & .462 & -.786 & .386 & .410 & -.222 &  .352 & -.772 & .355 & .359 & -.121 \\
			\thickhline
			5:0 & .505 & -.287 & .444 & -.141 & -.214 & .450 & -.152 & .479 & .767 & -.417 \\
			\hline
			5:1 & .166 & -.387 &  .243 & .238 & -.662 & .661 & -.110 & .217 & .531 & -.619 \\
			\hline
			5:2 & .018 & -.339 &  -.016 & .047 & -.447 & .441 & -.128 & .004 & .411 & -.675 \\
			\hline
			5:3 & .071 & -.233 &  .040 & -.027 & -.371 & .277 & -.127 & .056 & .296 & -.619 \\
			\thickhline
			6:0 & .300 & -.177 & .209 & -.027 & -.371 & .155 & -.147 & .301 & .191 & -.271\\
			\hline
			6:1 & -.016 & -.170 &  .043 & .112 & -.301 & .222 & .011 & .012 & .232 & -.221 \\
			\hline
			6:2 & -.094 & -.183 &  -.123 & -.069 & -.281 & .252 & .011 & -.148 & .272 & -.324 \\
			\hline
			6:3 & -.059 & -.131 & -.126 & -.153 & -.271 &  .263 & -.031 & -.147 & .209 & -.441 \\
			\thickhline
			7:0 & .297 & -.189 & .287 & .007 & -.220 & .015 & -.133 & .326 & .043 & -.208\\
			\hline
			7:1 & -.110 &  -.027 &  -.037 & .050 & -.354 & .216 & .052 & -.058 & .202 & -.224 \\
			\hline
			7:2 & -.140 & -.012 &  -.167 & -.058 & -.298 & .150 & .053 & -.158 & .148 & -.309 \\
			\hline
			7:3 & -.145 & .024 &  -.207 & -.100 & -.252 & .101 & .066 & -.201 & .104 & -.344 \\
			\thickhline
			8:0 & .256 & -.192 & .239 & .028 & -.131 & .003 & -.115 & .286 & .015 & -.045\\
			\hline
			8:1 & -.198 & .056 & -.124 & .009 & -.198 & .029 & .067 & -.158 & .056 & -.101 \\
			\hline
			8:2 & -.194 & .062 &  -.210 & -.061 & -.191 & .047 & .066 & -.223 & .057 & -.178 \\
			\hline
			8:3 & -.201 & .094 &  -.237 & -.092 & -.169 & .046 & .078 & -.258 & .040 & -.214 \\
			\hline
		\end{tabular}
		\caption{The Pearson product moment correlation between the probability of getting an optimal solution with $p$ iterations of QAOA on an $n$ vertex graph with the graph properties in columns two through eleven.}
		\label{tab:prob}
	\end{table}

	\begin{table}[H]\footnotesize
		\centering
		\begin{tabular}{|c|c|c|c|c|c|c|c|c|c|c|c|}
			\hline
			$n:p$ & Edges & Diam. & Clique num. & Bipartite & Eulerian  & Dist. reg. & Num. cut vertices & Num. min. odd cycles & Grp. size  & Num. orbits\\
			%\hline
			%3:0 &  &  &  &  &  &  &  &  &  & \\
			%\hline
			%3:1 &  1 & -1 & 1 & 1 & -1 & -1 & -1 & 1 & 1 & -1\\
			%\hline
			% 3:2 &  .981 & -.981 & .981 & .981 & -.981 & -.981 & -.981 & .981 & .981 & -.981\\
			\hline
			%3:2 & \rh{NEED} & \rh{NEED} & \rh{NEED} & \rh{NEED} & \rh{NEED} & \rh{NEED} & \rh{NEED} & N/A & \rh{NEED} & \rh{NEED}\\
			%\hline
			4:0 & 1 & -.812 & .908 & .781 & .070 & .552 & -.768 & .947 & .746 & -.417\\
			\hline
			4:1 & .983 & -.786 & .824 & .672 & -.101 & .669 & -.819 & .876 & .761 & -.548\\
			\hline
			4:2 & .799 & -.639 &  .479 & .355 & -.481 & .761 & -.920 & .583 & .608 & -.725 \\
			\hline
			4:3 & .783 & -.586 & .450 & .338 & -.448 &  .709 & -.901 & .579 & .572 & -.725 \\
			\thickhline
			5:0 & 1 & -.673 & .845 & .558 & -.247 & .267 & -.691 & .951 & .527 & -.424 \\
			\hline
			5:1 & .989 & -.671 &  .774 & .495 & -.255 & .305 & -.739 & .908 & .527 & -.464 \\
			\hline
			5:2 & .889 & -.641 &  .571 & .256 & -.118 & .180 & -.781 & .730 & .456 & -.463 \\
			\hline
			5:3 & .844 & -.564 &  .514 & .234 & -.087 & .072 & -.757 & .679 & .353 & -.369 \\
			\thickhline
			6:0 & 1 & -.673 & .768 & .466 & -.052 & .189 & -.684 & .933 & .323 & -.277 \\
			\hline
			6:1 & .991 & -.676 &  .697 & .401 & -.073 & .221 & -.729 & .886 & .319 & -.305 \\
			\hline
			6:2 & .926 & -.662 &  .544 & .250 & -.087 & .256 & -.750 & .749 & .312 & -.351 \\
			\hline
			6:3 & .875 & -.618 & .468 & .178 & -.098 &  .264 & -.747 & .676 & .274 & -.372 \\
			\thickhline
			7:0 & 1 & -.683 & .722 & .329 & -.079 & .056 & -.655 & .924 & .147 & -.225 \\
			\hline
			7:1 & .994 &  -.695 &  .663 & .295 & -.081 & .057 & -.698 & .884 & .142 & -.227 \\
			\hline
			7:2 & .953 & -.684 &  .558 & .214 & -.074 & .044 & -.711 & .790 & .132 & -.245 \\
			\hline
			7:3 & .916 & -.651 & .500 & .169 & -.067 & .030 & -.697 & .731 & .116 & -.252 \\
			\thickhline
			8:0 & 1 & -.691 & .682 & .222 & -.021 & .024 & -.603 & .913 & .050 & -.123\\
			\hline
			8:1 & .995 & -.707 & .629 & .203 & -.024 & .028 & -.646 & .876 & .048 & -.117 \\
			\hline
			8:2 & .967 & -.698 &  .550 & .158 & -.026 & .034 & -.652 & .803 & .046 & -.136 \\
			\hline
			8:3 & .936 & -.671 &  .507 & .125 & -.028 & .040 & -.640 & .752 & .042 & -.153 \\
			\hline
		\end{tabular}
		\caption{The Pearson product moment correlation between $\langle C\rangle_p$ for an $n$ vertex graph with the graph properties in columns two through eleven.}
		\label{tab:expectedvalue}
	\end{table}

	\begin{table}[H]\footnotesize
		\centering
		\begin{tabular}{|c|c|c|c|c|c|c|c|c|c|c|c|}
			\hline
			$n:p$ & Edges & Diam. & Clique num. & Bipartite & Eulerian  & Dist. reg. & Num. cut vertices & Num. min. odd cycles & Grp. size  & Num. orbits\\
			\hline
			%3:0 &  &  &  &  &  &  &  &  &  & \\
			%\hline
			%3:1 &  1 & -1 & 1 & 1 & -1 & -1 & -1 & 1 & 1 & -1\\
			%\hline
			%3:2 &  .981 & -.981 & .981 & .981 & -.981 & -.981 & -.981 & .981 & .981 & -.981\\
			%\hline
			%3:2 & \rh{NEED} & \rh{NEED} & \rh{NEED} & \rh{NEED} & \rh{NEED} & \rh{NEED} & \rh{NEED} & N/A & \rh{NEED} & \rh{NEED}\\
			%\hline
			4:0 & .857 & -.740 & .988 & .926 & .414 & .252 & -.446 & .926 & .602 & .017\\
			\hline
			4:1 & .635 & -.576 & .886 & .819 & .515 & .125 & -.133 & .753 & .491 & .206\\
			\hline
			4:2 & .747 & -.800 &  .552 & .440 & -.512 & .810 & -.882 & .525 & .630 & -.520 \\
			\hline
			4:3 & .466 & -.785 & .389 & .416 & -.219 &  .347 & -.777 & .360 & .356 & -.119 \\
			\thickhline
			5:0 & .770 & -.592 & .856 & .744 & -.414 & .389 & -.397 & .849 & .499 & -.421\\
			\hline
			5:1 & .428 & -.400 &  .569 & .599 & -.510 & .549 & -.164 & .546 & .421 & -.463 \\
			\hline
			5:2 & .125 & -.390 &  .152 & .091 & -.335 & .483 & -.154 & .147 & .424 & -.648 \\
			\hline
			5:3 & .138 & -.350 &  .103 & .060 & -.373 & .318 & -.206 & .116 & .329 & -.666 \\
			\thickhline
			6:0 & .720 & -.530 & .800 & .681 & -.040 & .061 & -.350 & .822 & .246 & -.116\\
			\hline
			6:1 & .374 & -.314 &  .539 & .541 & -.066 & .051 & -.103 & .515 & .192 & -.071 \\
			\hline
			6:2 & .166 & -.300 &  .218 & .234 & -.133 & .140 & -.050 & .193 & .258 & -.212 \\
			\hline
			6:3 & .042 & -.231 & -.035 & -.057 & -.256 &  .298 & -.099 & -.033 & .234 & -.420 \\
			\thickhline
			7:0 & .687 & -.489 & .727 & .494 & -.120 & .074 & -.337 & .794 & .127 &  -.158\\
			\hline
			7:1 & .315 &  -.258 &  .442 & .387 & -.140 & .112 & -.113 & .462 & .107 & -.116 \\
			\hline
			7:2 & .143 & -.210 &  .182 & .207 & -.164 & .134 & -.063 & .206 & .130 & -.202 \\
			\hline
			7:3 & .035 & -.128 &  -.015 & .028 & -.182 & .117 & -.023 & .010 & .112 & -.305 \\
			\thickhline
			8:0 &  .671 & -.469 & .666 & .347 & -.032 & -.007 & -.314 & .778 & .039 & -.034\\
			\hline
			8:1 & .299 & -.247 & .383 & .276 & -.042 & -.009 & -.129 & .443 & .031 & .024 \\
			\hline
			8:2 & .157 & -.197 &  .163 & .175 & -.063 & .004 & -.082 & .227 & .037 & -.015 \\
			\hline
			8:3 & .051 & -.115 &  -.002 & .043 & -.090 & .031 & -.037 & .043 & .039 & -.107 \\
			\hline
		\end{tabular}
		\caption{The Pearson product moment correlation between the level $p$ approximation ratio on an $n$ vertex graph with the graph properties in columns two through eleven.}
		\label{tab:gap}
	\end{table}
	
	\begin{table}[H]\footnotesize
		\centering
		\begin{tabular}{|c|c|c|c|c|c|c|c|c|c|c|c|}
			\hline
			$n:p$ & Edges & Diam. & Clique num. & Bipartite & Eulerian  & Dist. reg. & Num. cut vertices & Num. min. odd cycles & Grp. size  & Num. orbits\\
			\hline
			% 3:0 &   &  &  &  &  &  &  &  &  & \\
			%\hline
			%3:1 &  -1 & 1 & -1 & -1 & 1 & 1 & 1 & -1 & -1 & 1\\
			%\hline
			%3:2 &  1 & -1 & 1 & 1 & -1 & -1 & -1 & 1 & 1 & -1\\
			%\hline
			%3:2 & \rh{NEED} & \rh{NEED} & \rh{NEED} & \rh{NEED} & \rh{NEED} & \rh{NEED} & \rh{NEED} & N/A & \rh{NEED} & \rh{NEED}\\
			%\hline
			%4:0 &  -.965 & .793 & -.880 & -.842 & -.151 & .378 & .802 & -.941 & -.611 & .273\\
			%\hline
			4:1 & .295 & -.362 & .619 & .513 & .451 & .079 & .192 & .418 & .368 & .288\\
			\hline
			4:2 & .701 & -.766 &  .412 & .192 & -.567 & .897 & -.901 & .465 & .742 & -.809 \\
			\hline
			4:3 & .362 & -.979 & .483 & .483 &  NA & NA  & -.682 & .362 & .511 & .422 \\
			\thickhline
			% 5:0 &  -.856 & .533 & -.787 & -.502 & -.033 & -.153 & .540 & -.814 & -.302 & .066\\
			%\hline
			5:1 & .175 & -.276  & .305 & .306  & -.612 & .645  & -.001 & .271  & .487 & -.495 \\
			\hline
			5:2 & .015 & -.347  & -.045 & -.140  & -.490 & .571  & -.140 & -.011  & .507 & -.710 \\
			\hline
			5:3 & .204 & -.251  & .177 & .095  & -.423 & NA  & -.114 & .158  & .700 & -.623 \\
			\thickhline
			%6:0 &  -.890 & .586 & -.773 & -.513 & -.030 & .042 & .571 & -.862 & -.191 & .125 \\
			%\hline
			6:1 & -.045 & -.053  & .139 & .209  & -.110 & .112  & .149 & .077  & .200 & -.078 \\
			\hline
			6:2 & -.153 & -.124  & -.240 & -.203  & -.228 & .310  & .045 & -.252  & .395 & -.318 \\
			\hline
			6:3 & -.003 & -.152  & -.139 & -.206  & -.393 & .438  & -.099 & -.119  & .337 & -.545 \\
			\thickhline
			%7:0 &  -.885 & .570 & -.745 & -.344 & .003 & .011 & .531 & -.721 & -.066 & .099 \\
			%\hline
			7:1 & -.157 & .045  & .018 & .122  & -.182 & .180  & .148 & -.025  & .160 & -.090 \\
			\hline
			7:2 & -.235 & .012  & -.324 & -.146  & -.244 & .226  & .080 & -.315  & .235 & -.237 \\
			\hline
			7:3 & -.114 & .029  & -.210 & -.159  & -.308 & .211  & .069 & -.208  & .176 & -.392 \\
			\thickhline
			%8:0 &  -.893 & .556 & -.730 & -.243 & -.007 & .000 & .438 & -.861 & -.027 & .115\\
			%\hline
			8:1 & -.237 & .090  & -.078 & .078  & -.057 & .007  & .124 & -.104  & .041 & .053 \\
			\hline
			8:2 & -.259 & .048  & -.389 & -.091  & -.102 & .045  & .077 & -.361  & .079 & -.077 \\
			\hline
			8:3 & -.187 & .107  & -.279 & -.159  & -.166 & .082  & .088 & -.302  & .083 & -.226 \\
			\hline
		\end{tabular}
		\caption{The Pearson product moment correlation between the $\Delta$ ratio at $p$ on an $n$ vertex graph with the graph properties in columns two through eleven.}
		\label{tab:changeingap}
	\end{table}

	\begin{table}[H]\footnotesize
		\centering
		\begin{tabular}{|c|c|c|c|c|c|c|c|c|}
			\hline
			$n:p$ & B. $P(C_{max})$ & N.B. $P(C_{max})$ & B. $\langle C \rangle$ & N.B.$\langle C \rangle$ & B. Level $p$ A.R. & N.B. Level $p$ A.R. & B. $\Delta$ ratio $p-1$ to $p$ & N.B. $\Delta$ ratio $p-1$ to $p$ \\
			\hline
			4:0 & .125 & .0625 & 1.667 & 2.5 & .5 & .681 & NA & NA \\
			\hline
			4:1 & .481 & .602 & 2.566 & 3.216 & .772 & .879 & .544 & .634 \\
			\hline
			4:2 & .889 & .928 & 3.180 & 3.586 & .949 & .978 & .762 & .825 \\
			\hline
			4:3 & .993 & .999 & 3.326 & 3.666 & .998 & 1.000  & .973 & .994 \\
			\thickhline
			5:0 & .0625 & .049 & 2.3 & 3.344 & .5 & .658 & NA & NA \\
			\hline
			5:1 & .368 & .495 & 3.436 & 4.323 & .750 & .857 & .500 & .605 \\
			\hline
			5:2 & .746 & .725 &  4.222 & 4.685 & .918 & .928 & .661 & .587 \\
			\hline
			5:3 & .907 & .900 &  4.470 & 4.926 & .970 & .974 & .731 & .744 \\
			\thickhline
			6:0 & .031 & .028 & 3.118 & 4.447 & 0.500 & 0.644 & NA & NA \\
			\hline
			6:1 & .260 & .311 &  4.542 & 5.672 & .733 & .826 & .465 & .522 \\
			\hline
			6:2 & .586 & .549 &  5.449 & 6.179 & .873 & .900 & .519 & .442 \\
			\hline
			6:3 & .818 & .744 & 5.949 & 6.506 & .951 &  .946 & .620 & .511 \\
			\thickhline
			7:0 & .016 & .016 & 3.875 & 5.693 & .067 & .074 & NA & NA \\
			\hline
			7:1 & .182 &  .213 &  5.554 & 7.187 & .438 & .482 & .182 & .213 \\
			\hline
			7:2 & .469 & .424 &  6.598 & 7.827 & .851 & .886 & .464 & .396 \\
			\hline
			7:3 & .691 & .605 & 7.201 & 8.225 & .927 & .930 & .519 & .409 \\
			\thickhline
			8:0 & .008 & .011 & 4.797 & 7.246 & .5 & .646 & NA & NA \\
			\hline
			8:1 & .133 & .139 & 6.773 & 9.022 & .710 & .808 & .420 & .462 \\
			\hline
			8:2 & .385 & .317 & 7.983 & 9.801 & .832 & .877 & .420 & .367 \\
			\hline
			8:3 & .606 & .482 &  8.762 & 10.273 & .911 & .920 & .467 & .350 \\
			\hline
		\end{tabular}
		\caption{The average value of bipartite graphs versus non-bipartite graphs, where the columns beginning with ``B." refer to bipartite graphs, and those with ``N.B." refer to non-bipartite graphs. The abbreviation ``A.R.' stands for the approximation ratio. Three vertex graphs are excluded since there is one bipartite three vertex graph and one non-bipartite.}
		\label{tab:bipartite}
	\end{table}
	
	\begin{table}[H]\footnotesize
		\centering
		\begin{tabular}{|c|c|c|c|c|c|c|c|c|}
			\hline
			$n:p$ & E. $P(C_{max})$ & N.E. $P(C_{max})$ & E. $\langle C \rangle$ & N.E.$\langle C \rangle$ & E. Level $p$ A.R. & N.E. Level $p$ A.R. & E. $\Delta$ ratio $p-1$ to $p$ & N.E. $\Delta$ ratio $p-1$ to $p$ \\
			\hline
			4:0 & .125 & .088 & 2 & 2.1 & .5 & .608 & NA & NA \\
			\hline
			4:1 & .531 & .543 & 3.000 & 2.869 & .75 & .841 & .5 & .607 \\
			\hline
			4:2 & 1 & .890 & 4 & 3.260 & 1 & .956 & 1 & .752 \\
			\hline
			4:3 & 1 & .995 & 4 & 3.395 & 1 & .998  & 1 & .980 \\
			\thickhline
			5:0 & .070 & .048 & 3.5 & 3 & .698 & .602 & NA & NA \\
			\hline
			5:1 & .775 & .392 & 4.513 & 4.017 & .912 & .813 & .764 & .537 \\
			\hline
			5:2 & .913 & .687 & 4.762 & 4.531 & .960 & .918 & .831 & .551 \\
			\hline
			5:3 & .990 & .881 & 4.965 & 4.782 & .994 & .968 & .963 & .689 \\
			\thickhline
			6:0 & .078 & .025 & 4.438 & 4.231 & .633 & .621 & NA & NA \\
			\hline
			6:1 & .481 & .290 &  5.766 & 5.480 & .827 & .811 & .552 & .510 \\
			\hline
			6:2 & .749 & .539 &  6.397 & 6.043 & .915 & .894 & .566 & .445 \\
			\hline
			6:3 & .925 & .742 & 6.816 & 6.391 & .976 &  .945 & .796 & .507 \\
			\thickhline
			7:0 & .041 & .015 & 6.054 & 5.578 & .075 & .074 & NA & NA \\
			\hline
			7:1 & .431 &  .201 &  7.570 & 7.081 & .843 & .807 & .569 & .475 \\
			\hline
			7:2 & .666 & .415 &  8.203 & 7.744 & .912 & .883 & .518 & .394 \\
			\hline
			7:3 & .833 & .600 & 8.597 & 8.153 & .955 & .929 & .609 & .406 \\
			\thickhline
			8:0 & .027 & .011 &7.438  & 7.202 & .656 & .643 & NA & NA \\
			\hline
			8:1 & .276 & .137 & 9.245 & 8.981 & .820 & .806 & .491 & .461 \\
			\hline
			8:2 & .525 & .315 &  10.067 & 9.766 & .893 & .876 & .426 & .366 \\
			\hline
			8:3 & .708 & .480 &  10.580 & 10.242 & .937 & .919 & .472 & .350 \\
			\hline
		\end{tabular}
		\caption{The average value of Eulerian graphs versus non-Eulerian graphs, where the columns beginning with ``E." refer to Eulerian graphs, and those with ``N.E." refer to non-Eulerian graphs. The abbreviation ``A.R." stands for the approximation ratio. Three vertex graphs are excluded since there is one Eulerian three vertex graph and one non-Eulerian. The entries for Eulerian $4:3$ are are the same as Eulerian $4:2$ since the only Eulerian graph on four vertices achieves its maximum values after two iterations.}
		\label{tab:eulerian}
	\end{table}
	
	\twocolumngrid
	\bibliographystyle{unsrt}
	\bibliography{references.bib}
\end{document}